\documentclass[twocolumn,a4paper,showpacs,preprintnumbers,amsmath,amssymb,prl]{revtex4}

\usepackage{graphicx} 
\usepackage{dcolumn}
\usepackage{bm} 
\usepackage{color}

\begin{document}

\hyphenation{te-tra-go-nal}

\bibliographystyle{apsrev}

\title{Direct observation of room temperature high-energy resonant excitonic effects in graphene}

\author{I. Santoso$^{1,3,4,8}$, P.K Gogoi$^{1,2}$, H.B. Su$^{5}$, H. Huang$^{2}$, Y. Lu$^{2}$, D. Qi$^{1,2,3}$, W. Chen$^{2,4}$, M.A. Majidi$^{1,2}$, Y. P. Feng$^{1,2}$, A. T. S. Wee $^{1,2}$, K. P. Loh$^{1,3}$, T. Venkatesan$^{1,2}$, R. P. Saichu$^{6}$, A. Goos$^{6}$, A. Kotlov$^{7}$, M. R\"{u}bhausen$^{6,1}$, A. Rusydi$^{1,2,3,6}$ }
\email{phyandri@nus.edu.sg}
\affiliation{$^{1}$NUSNNI-Nanocore, National University of Singapore, Singapore 117576}
\affiliation{$^{2}$Department of Physics, National University of Singapore, Singapore 117542}
\affiliation{$^{3}$Singapore Synchrotron Light Source, National University of Singapore, 5 Research Link, Singapore 117603, Singapore}
\affiliation{$^{4}$Department of Chemistry, National University of Singapore, Singapore 117543}
\affiliation{$^{5}$Division of Material Science, Nanyang Technical University, 50 Nanyang Avenue Singapore 639798}
\affiliation{$^{6}$Institut f\"{u}r Angewandte Physik, Universit\"{a}t Hamburg, Jungiusstrasse 11, 20355 Hamburg, Germany. Center for Free Electron Laser Science (CFEL), D-22607 Hamburg, Germany}
\affiliation{$^{7}$Hamburger Synchrotronstrahlungslabor (HASYLAB) at Deutsches Elektronen-Synchrotron (DESY), Notkestra$\beta$e 85, 22603 Hamburg, Germany}
\affiliation{$^{8}$Jurusan Fisika, Universitas Gadjah Mada, BLS 21 Jogyakarta, Indonesia}

\date{\today}

\begin{abstract}

Using a combination of ultraviolet-vacuum ultraviolet reflectivity and spectroscopic ellipsometry, we observe a resonant exciton at an unusually high energy of 6.3eV in epitaxial graphene. Surprisingly, the resonant exciton occurs at room temperature and for a very large number of graphene layers $N$$\approx$75, thus suggesting a poor screening in graphene. The optical conductivity ($\sigma_1$) of resonant exciton scales linearly with number of graphene layer (up to \emph{at least} 8 layers) implying quantum character of electrons in graphene. Furthermore, a prominent excitation at 5.4eV, which is a mixture of interband transitions from   $\pi$ to $\pi^{*}$ at the M point and a $\pi$ plasmonic excitation, is observed. In contrast, for graphite the resonant exciton is not observable but strong interband transitions are seen instead. Supported by theoretical calculations, for $N \leq$ 28 the $\sigma_1$ is dominated by the resonant exciton, while for $N >$ 28 it is a mixture between exitonic and interband transitions. The latter is characteristic for graphite, indicating a crossover in the electronic structure. Our study shows that important elementary excitations in graphene occur at high binding energies and elucidate the differences in the way electrons interact in graphene and graphite.
\\

\end{abstract}

\maketitle

Graphene, a one layer thick carbon honeycomb structure, has recently attracted a lot of attention due its exotic quantum properties\cite{reviews}. In addition to the well-known electronic properties, such as ballistic electron transport\cite{reviews,NovScience01,NovNature01}, quantum Hall effect\cite{ZhangNature01}, tunable band gap\cite{ohtaScience,zhangNature02}, and physics driven by many-body interactions\cite{BostwickNatPhys}, graphene also displays very interesting optical properties. For instance, due to its low-energy excitation graphene becomes highly transparent in the visible spectral range  and its infrared optical conductivity can be tuned using a gate voltage\cite{LiNatPhys}. These particular properties of graphene potentially result in exciting optoelectronic applications.

\begin{figure}
\begin{center}
\includegraphics[width=3.2in]{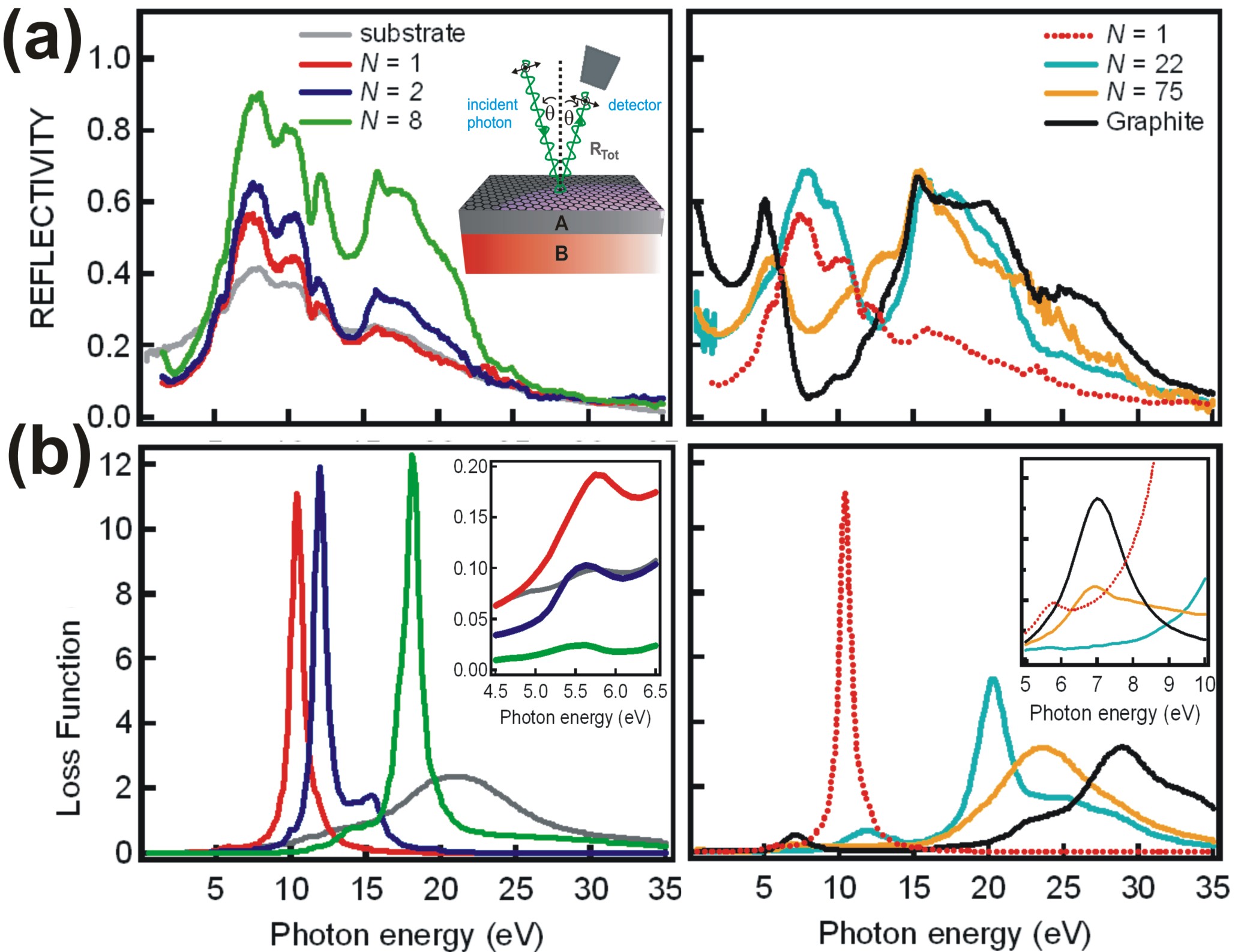}
\caption{\label{fig:fig1-R} (color online) Room temperature experimental results of (a) reflectivity (b) loss function, Im ($\varepsilon^{-1}$).
The inset of (a) shows the experimental geometry while the inset of (b) shows ($\varepsilon^{-1}$) on an expanded scale from 4.5 to 6.5 eV and from 5 to 10 eV.}
\end{center}
\end{figure}

Recent theoretical studies based on the ab-initio GW and Bethe-Salpeter equation (BSE) approach by Yang et al.\cite{YangPRL}  and Trevisanutto et al.\cite{TrevisPRB} have predicted the existence and have highlighted the importance of resonant excitonic effects in the optical absorption of graphene. However, there is disagreement as to the origin and position of the exciton. In Ref. \cite{YangPRL} the calculations were done up to 7 eV and the resonant exciton predicted to occur at 4.6 eV due to electron-hole interaction in
$\pi^{*}$- and $\pi$-band at the M point. While in Ref. \cite{TrevisPRB} the calculations were done at a much higher energy of 22 eV and the resonant exciton predicted to appear at 8.3 eV due the background single particle continuum of dipole forbidden transition at the the $\Gamma$ point. Despite their disagreement, both have agreed that resonant exciton plays important role for elementary excitations in graphene and thus its understanding is crucial.

A direct way to probe resonant exciton in graphene is to measure its complex dielectric response in wide energy range\cite{TrevisPRB}. Despite reports on the optical properties of graphene over the visible energy range\cite{BostwickNatPhys,NairScience,KuzmenkoPRL01,MakPRL01,LatilPRL,OhtaPRL,KravetsPRB} there has been no reliable optical conductivity
data at higher energies. Therefore, it is crucial to study the optical properties of graphene in an unprecedented high energy range not only to provide a direct evidence of the resonant excitonic effects \emph{per se}, but also for our understanding of electronic structure of graphene.

Herewith, we report the optical conductivity ($\sigma_{1}$) of epitaxial graphene on 6H-SiC(0001)/buffer layer substrate in an unprecedented  wide photon energy range from 0.5 to 35 eV using a combination of spectroscopic ellipsometry and UV-VUV reflectrometry \cite{RusydiPRB01}.
It has been recently shown that the combination of these two techniques enables a stabilized Kramers-Kronig transformation\cite{RusydiPRB01} which is crucial to resolve precisely the dielectric function. We study the evolution of $\sigma_{1}$ as a function of the number of layers \emph{N} (henceforth $\sigma_{1,N}$) including graphite.

\begin{figure}
\begin{center}
\includegraphics[width=3.4in]{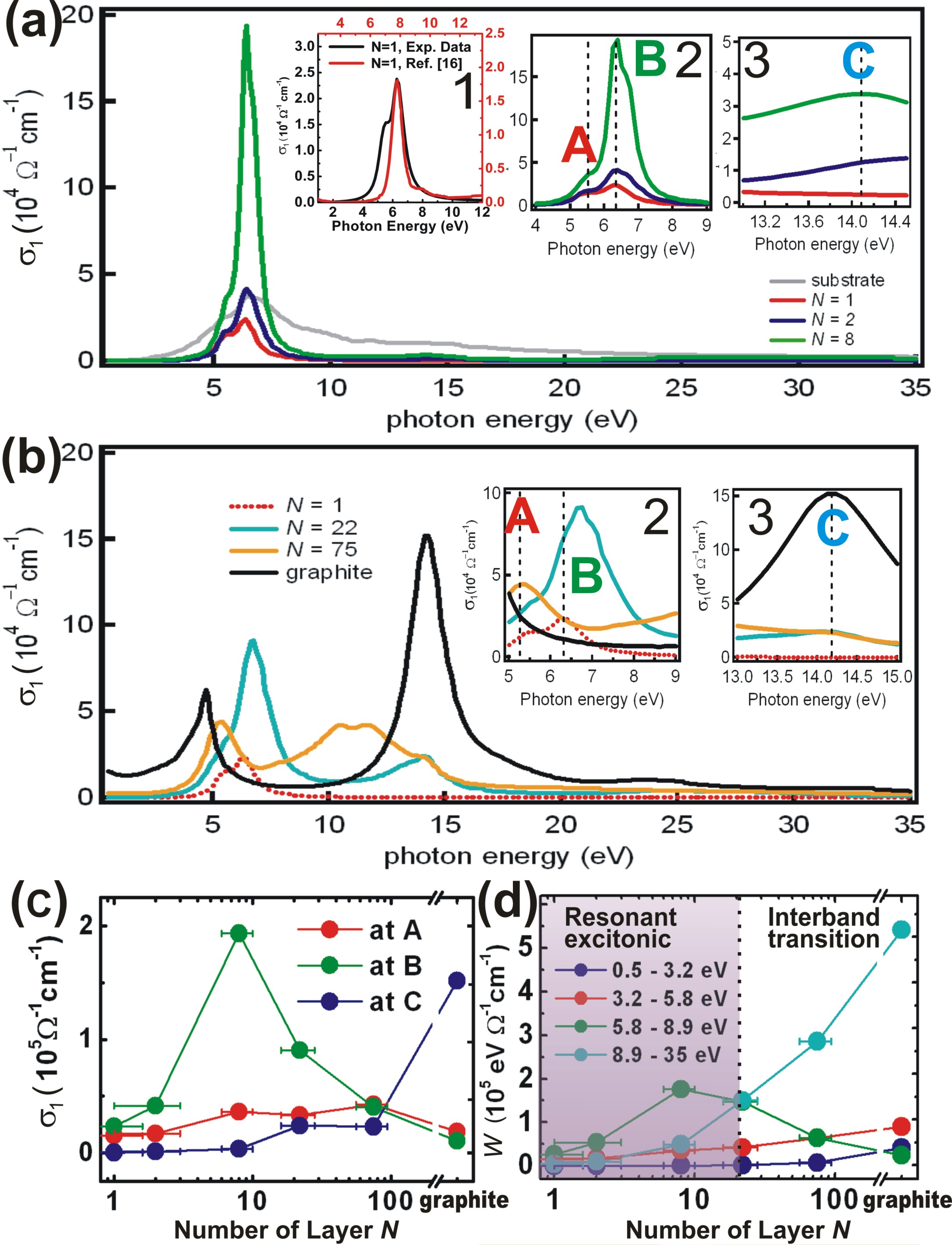}
\caption{\label{fig:fig2-sigma1} (color online) The optical conductivity ($\sigma_{1}$) of (a) substrate and graphene ($N = 1, 2, 8$) and (b) graphene ($N = 22, 75$) and graphite show three peaks at 5.4 eV (label A), 6.3 eV (label B) and 14.1 eV (label C), respectively.
Inset 1 shows the comparison of $\sigma_{1,1}$ between experimental data and theoretical calculations.
Insets 2 and 3 show the $\sigma_{1}$ on an expanded scale at various energy ranges. (c) The value of $\sigma_1$ at A, B and C as a function of $N$. (d) The partial spectral weight (\emph{W}) as a function of $N$.}
\end{center}
\end{figure}

Epitaxial graphene samples on 6H-SiC(0001) were prepared by Si desorption process in Si-flux as described in previous studies\cite{WeiSS,HanACS}. This process was found to lead to multilayer graphene with ordered stacking\cite{HanACS}. The thickness was determined using scanning tunneling microscopy and Raman spectroscopy\cite{HanACS,ShivaJEM}. For reference, we have used a cleaved highly ordered pyrolytic graphite. The reproducibility was checked by fabricating and repeating individual measurements on three different samples that were grown under identical conditions at least 3 different locations on each sample yielding reproducible results. The details of the samples are described in the supplementary section.

Figure~\ref{fig:fig1-R}(a) shows room temperature reflectivity data for \emph{N}-dependent epitaxial graphene, graphite as well as the substrate. Interestingly, one can see that the reflectivity of epitaxial graphene is very much dependent on \emph{N} and very much distinguished from that of graphite. The reflectivity has rich and distinct structures especially in the range from 5 to 8 eV for \emph{N} = 1, 2, and 8 while additional structures occur around 14 to 20 eV for \emph{N} = 22, and 75. On the other hand, the reflectivity of graphite has shown strong structures below 0.5 eV, $\sim5$ eV, and around 14 to 20 eV which are similar to published result\cite{TaftPR}. Further rigorous discussion will be achieved by analysing the $\sigma_1$. (Details of analysis of reflectivity, complex dielectric function and optical conductivity are shown in the Supplementary.)

The $\sigma_{1,N}$ shows striking results (Figs.~\ref{fig:fig2-sigma1} (a) and (b)). For $N = 1$, the $\sigma_{1,1}$ is dominated by two well-defined peaks, very pronounced at 6.3 eV (peak B) and less pronounced at 5.4 eV (peak A), with almost equal full-width at half maximum (FWHM) of $\thicksim$0.8 eV. These structures, especially the peak B, are considerably sharp for this high energy range.

Now, we shall focus our discussion on peak B as it shows an interesting dependence on $N$ (Fig.~\ref{fig:fig2-sigma1}(c)). The $\sigma_{1, N}(B)$ increases rapidly towards $N$=8, while its FWHM and peak position remain independent of $N$. For $N > 8$, $\sigma_{1, N}$(B) decreases while the peak position shifts towards higher energy (6.7 eV and 11 eV for N = 22 and N = 75, respectively). For graphite, peak B disappears and our result is similar to the published data\cite{TaftPR}.

To find out the origin of peak B, we have directly compared our experimental data with theoretical calculations\cite{YangPRL,TrevisPRB}. As shown in inset 1 of Fig.~\ref{fig:fig2-sigma1}(a), the line shape and the  $\sigma_{1, 1}$(B) between experimental data and theoretical calculations from Ref. \cite{TrevisPRB} are surprisingly very similar. This comparison shows a decisive evidence of the high energy resonant exciton in epitaxial graphene. The resonant exciton arises from dipole transitions of the single-particle continuum. However, the observed resonant exciton peak occurs \emph{red shifted} compared to the calculations. Our calculations show that the red shifted exciton could result from film-substrate interactions, which are not included in the previous calculations (see below for the details). Thus, the origin of peak B is the high energy resonant exciton as predicted.

Furthermore, our detail study rules out the interband transitions and plasmonic excitations as origin of peak B. Based on Density Functional Theory (DFT), we have calculated optical conductivity for $N$ = 1 and 2, and found that the peak B does not originate from interband transitions (see Fig. 3a and discussion below). Secondly, one may argue that we have to consider contributions from a plasmonic excitation. To address this issue, we have studied in detail an energy loss function Im$(\varepsilon^{-1})$ which can reveal collective excitations such as plasmonic excitations\cite{MarinPRL01,EberPRB}. As shown in Fig. 1(b), there is no feature that can be attributed to a plasmonic excitation. For $N$ = 1, Im$(\varepsilon^{-1})$ is dominated by a strong structure at $\sim$10 eV and a weak structure at ~5.5 eV. These structures are from $\pi+\sigma$ and $\pi$ in-plane plasmon modes, respectively\cite{EberPRB}. Based on the symmetry, the 5.5 eV plasmon peak structure is only visible for light polarization parallel to \emph{c}-axis (\textbf{E}$\|c$). In our measurement, the {\bf E} was mixed between {\bf E}$\|c$ and {\bf E}$\bot c$ with main contribution from {\bf E}$\bot c$, thus the observed $\pi$ plasmon is very weak. On the other hand, $\pi+\sigma$ plasmon mode is very strong in our geometry which is consistent with EELS measurements\cite{EberPRB,JiongPRB}. As \emph{N} increases, the $\pi+\sigma$ plasmon mode shifts toward higher energy and gets broader, while the structure at 5.5 eV is nearly \emph{N}-dependent. This {\it blueshift} may be due to strong effects of the interlayer Coulomb interaction on the total plasmon\cite{MarinPRL01}.

\begin{figure}
\begin{center}
\includegraphics[width=3.4in]{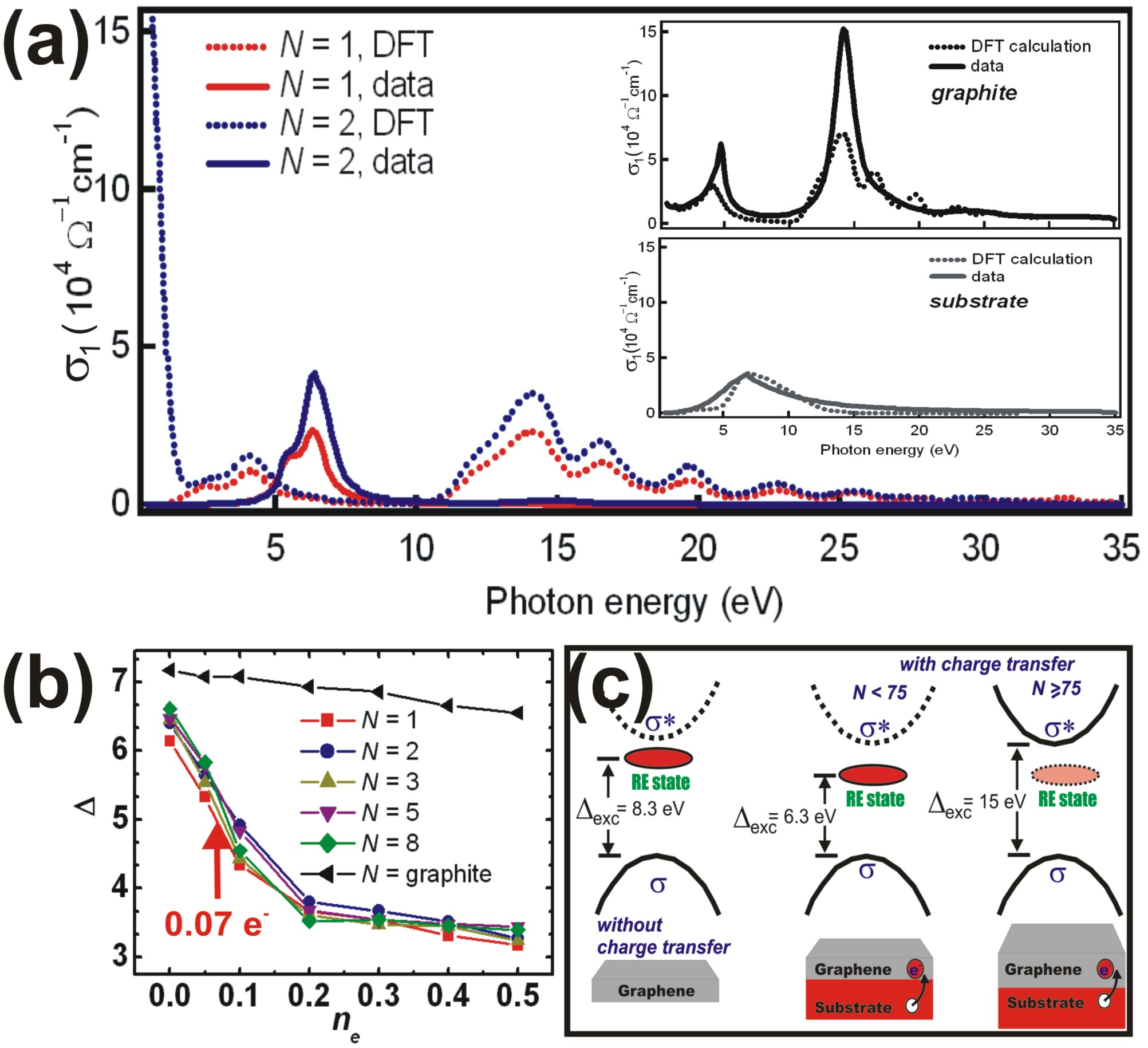}
\caption{\label{fig:fig3-model} (color online) (a) comparison between experimental data and calculated $\sigma_{1}$. {\bf Upper inset :} $\sigma_{1}$ spectra of graphite. {\bf Lower inset :} $\sigma_{1}$ spectra of the substrate. (b) The background single-particle continuum transitions ($\Delta$) at the $\Gamma$ point as a function of nominal charge transfer ($n_{e}$) from substrate to graphene for various graphene layer ($N$). Red arrow shows the observed charge transfer in our sample. (c) A proposed model of optical absorption of exciton in graphene and graphene on substrate as function of $N$. RE stands for the resonant exciton and ($\Delta_{exc}$) is excitation energy.}
\end{center}
\end{figure}

The exciton energy of 6.3 eV suggests an untypically high binding energy which can also be made responsible for the presence of the exciton up to room temperature. If one considers the angular-resolved photoemission data, the distance of  $\sigma$ band at $\Gamma$ point to Fermi level is around 4.5 eV\cite{EmtsevPRB}. By assuming the distance of $\sigma^*$ band to Fermi level to be similar to that of $\sigma$  band one will end up with 9 eV as the separation between $\sigma$ and $\sigma^*$ bands. Based on these values the binding energy of the resonant exciton is 2.7 eV which is 108 times larger than the thermal energy. This binding energy is significantly larger as compared to that of ZnO which is around 60 meV (~2.4 times thermal energy)\cite{Thomas60} and is very clearly seen at room temperature while the binding energy of the excition for GaAs is about 3.4 meV and can only be seen at low temperatures\cite{Hegarty85}. Such a very high binding energy exciton observed on graphene is indeed remarkable and a direct consequence of its low dimensionality in conjunction with its intriguing electronic properties.

We next discuss peak A and its origin. As shown in Fig.~\ref{fig:fig2-sigma1}(c),  $\sigma_{1,N}(A)$ increases monotonically as a function of $N$. Interestingly, for $N > 8$ as well as for graphite the position of peak A shifts towards lower energies as opposed to peak B, while the line shape is rather symmetric and the FWHM is broader for larger $N$. Based on electronic band structure calculations\cite{LatilPRL}, peak A is the result of interband transitions from the $\pi$ to $\pi^{*}$ bands at M-point where the van Hove singularity occurs. Furthermore, from analysis of energy loss function (Fig. \ref{fig:fig1-R}(b)) the Im$(\varepsilon^{-1})$ shows a weak  $\pi$-plasmon contribution. Thus one can conclude that the peak A is a mixture between interband transitions and a plasmonic excitation.

Another notable observation is a broad structure at 14.1 eV (peak C). The peak C is absent for $N = 1$ and $\sigma_{1,N}$(C) increases dramatically for $N \geq 2$ (Fig.~\ref{fig:fig2-sigma1}(c)). Interestingly, for graphite the $\sigma_{1}$ is dominated by peak C with symmetric lineshape. Based on electronic band structure calculations\cite{TrevisPRB,LatilPRL,AhujaPRB}
this structure is interband transitions from $\sigma$ to $\sigma^{*}$ bands at $\Gamma$ point. Peak C clearly has an intimate relationship with optical conductivity in graphite while peak B is a unique characteristic of graphene.

We next discussed partial spectral weight integral ($W$) because it describes the effective number of electrons excited by photons of respective energy. The $\sigma_{1}$ is restricted by the f-sum rule: $\int^{\infty}_{0}\sigma_{1}(E)dE=\frac{\pi n e^{2}}{2m^*}$, where $n$ is the electron density, $e$ is elementary charge and $m^*$ is effective electron mass. Hence, one can extract the $W\equiv\int^{E_{2}}_{E_{1}}\sigma_{1}(E)dE$,
for various energy ranges. Due to the f-sum rule and charge conservation, the $W$ is constant, and thus one can study the spectral weight transfer and reveal interactions as well as the effect of $N$ in the broad energy range of 0.5 to 35 eV.
Figure ~\ref{fig:fig2-sigma1}(d) shows $W$ for the different energy regions: 0.5 to 3.2 eV (region I, $W_{I}$), 3.2 to 5.8 eV (region II, $W_{II}$), 5.8 to 8.9 eV (region III, $W_{III}$), and 8.9 to 35.0 eV (region IV, $W_{IV}$).

The $W_{I}$ is mainly governed by the transition around Dirac cone, i.e., $\pi$ to $\pi^{*}$ around $K$ point in Brillouin zone which is consistent with previous publications\cite{BostwickNatPhys,MakPRL01}. While the $W_{II}$, $W_{III}$, and $W_{IV}$ show the main contribution from the peak A, B, and C, respectively. Interestingly, $W_{I}$ and  $W_{II}$ show almost similar trend in which the $W$ increases monotonically as function of $N$, while the $W_{III}$ and $W_{IV}$ show  completely different fashion. For $N \leq 8$ the $W_{III}$ increases as $N$ increases while for $N > 8$ $W_{III}$ decreases as $N$ increases. On the other hand, $W_{IV}$ increases rapidly for $N \geq 8$. In graphite, $W_{IV}$ is maximum while $W_{III}$ obtains its smallest value. We find a crossover between $W_{III}$ and $W_{IV}$ at $N \sim 28$ (see Fig. ~\ref{fig:fig2-sigma1}(d)). Thus, we propose that for $N < 28$ the optical conductivity of epitaxial multilayer graphene is dominated by high energy resonant excitonic effects, while for $N > 28$, the optical conductivity is dominated by interband transitions.

To gain further insight, we have calculated the optical conductivity ($\sigma_{1,calc}$) of graphene, graphite, and substrate using Density Functional Theory (DFT). We have studied the role of interband transitions and have compared them with experimental results (see Fig.~\ref{fig:fig3-model}(a)). Noting that our calculations do not include electron-hole ($e-h$) interactions, thus  ($\sigma_{1,calc}$) is mainly driven by interband transitions and serves as reference for the non-correlated case. (The detail of calculation is given in supplementary section.) Based on the DFT calculation,  ($\sigma_{1,calc}$) mimics reasonable well the  ($\sigma_{1}$) for graphite as well as for 6H-SiC(0001)/buffer layer. For graphite, the peak at ~4 eV is dominated by transitions from $\pi\rightarrow\pi^{*}$ bands while the peak at ~14.1 eV is transitions from $\sigma\rightarrow\sigma^{*}$ bands. This is consistent with previous theoretical study\cite{AhujaPRB}.

On the other hand, the DFT calculations for graphene show that its optical conductivity cannot be explained with interband transitions. As shown in Figure ~\ref{fig:fig3-model}(a), the calculated optical conductivity for graphene shows a completely different result than the experimental data and thus our attempt to mimic $\sigma_{1}$ for $N$ = 1 and 2 failed. Calculations based on DFT show that the  $\sigma_{1,calc}$ for graphene ($N$ = 1, 2) above 3 eV is very similar to graphite. This is in contrast to our experimental results. This further supports that the  $\sigma_{1,N}$ is mainly driven by strong $e-h$ interactions which lead to high energy resonant exciton and form unusual electronic band structure while the optical conductivity of graphite is mainly driven by interband transitions. It is shown that for $N$ = 1, once one turns on the $e-h$ interactions, the peak around 4 eV and peak around 14.1 eV (inset of Fig.~\ref{fig:fig3-model}(a)) vanish resulting new and very strong peak in between\cite{TrevisPRB}.

To find out the origin of the redshift of the excitonic excitation, we have calculated the variation of the $\sigma$ bands and the background single-particle continuum transitions ($\Delta$) as function of charge transfer ($n_{e}$) and strain. We find out that while the $\sigma$ bands are nearly independent from these two effects, the background single-particle continuum transitions are very much dependending on the charge transfer $n_{e}$ from substrate to the graphene layer (Fig. ~\ref{fig:fig3-model}(b)). Based on our DFT calculations, the charge transfer from substrate to graphene $n_{e}$ is about $\sim$0.07e (per graphene unit) and thus the background single-particle continuum transitions reduce by $\sim$1.4 eV compared to that of for $n_{e} = 0$. The strain in graphene due to the lattice mismatch with the substrate reduce the background single-particle continuum transitions by $~0.6$ eV. These two effects altogether decrease the single-particle continuum transitions to about 2.0 eV in total. Thus, one may expect to see the exciton at lower energy of $\sim$6.3 eV. Our result may suggest that energy excitation of exciton depends on the $n_{e}$ and the strain.

Interestingly, our calculation shows that $\Delta$ does not dependent on $N$ significantly (Fig.~\ref{fig:fig3-model}(b)). This is in fact consistent with our experimental result. In contrast, the theoretical predictions in Ref.\cite{TrevisPRB} show an energy shift between the exciton in graphene and bilayer graphene which is attributed to enhanced screening. Our experimental results show, however, that the position of the exciton does not depend on the $N$ value till about to $N$ = 22. This can be reconciled with the Ref.\cite{TrevisPRB} only if one assumes that screening effects are significantly reduced. Thus, by comparing the experimental results and theoretical calculations, one may conclude that screening effects in graphene are much weaker than one would expect. Altogether our results suggest that graphene belongs to the class of strongly correlated electron systems.

Finally, based on our experimental results of high energy optical conductivity, DFT  calculations, and recent {\it ab initio} GW-BSE calculations \cite{TrevisPRB}, we have proposed the following phase diagram (Fig.~\ref{fig:fig2-sigma1}(c-d)) and a model for optical absorption which is proportional to $\sigma_1$ (Fig. ~\ref{fig:fig3-model}(c)). For free standing graphene of $N$=1, the optical absorption is dominated by high energy resonant exciton effects which occur around 8.3 eV. However, in our case due to the charge transfer and lattice mismatch between graphene and the substrate, the resonant exciton reduces to 6.3 eV. For $N < 28$, the optical absorption shows very strong resonant excitonic-like structure  and weak interband transitions. While for $28 < N < 75$ the optical absorption shows a mixture between exciton and interband transitions (Figs.~\ref{fig:fig2-sigma1}(d) and ~\ref{fig:fig3-model}(c)). For $N > 75$ as well as for graphite, the optical absorption is dominated by interband transitions.

In summary, we have observed a high energy room temperature stable resonant exciton at $\sim$6.3eV with a large binding energy of $\sim$2.7 eV in the optical conductivity of multilayer epitaxial graphene revealing strong collective $e-h$ interactions. The resonant exciton persists for very large $N$ and thus dominates the electronic properties owing to the poor screening in graphene. Furthermore, the mixture of interband transition from $\pi$ to $\pi^{*}$ at the M point and weak $\pi$ plasmonic excitation gives rise to peak in the optical conductivity at 5.4 eV. These findings demonstrate the importance of high energy optical conductivity and have strong implications on the understanding of the electronic structure of epitaxial graphene.

We are grateful to George A. Sawatzky, Antonio Castro Neto, and Alexey Kuzmenko for their very useful discussions. This work is supported by NRF-CRP grant Tailoring Oxide Electronics by Atomic Control, NRF-CRP grant graphene and Related Materials R-143-000-360-281, NRF-CRP grant Multi-functional spintronic materials and devices, NUS YIA, Advance Material (NanoCore) R-263-000-432-646, BMBF 05KS7GUABMBF07-204 and MOE-AcRF-Tier-1 (Grant No. M52070060).

\end{document}